\documentclass[twocolumn,twocolappendix]{aastex63}

\usepackage{xspace}

\newcommand{\Ha}{H$\alpha$\xspace}
\newcommand{\Hb}{H$\beta$\xspace}
\newcommand{\NII}{[N{\sc ii}]\xspace}
\newcommand{\SII}{[S{\sc ii}]\xspace}
\newcommand{\OIII}{[O{\sc iii}]\xspace}
\newcommand{\gleam}{\textsc{gleam}\xspace}
\newcommand{\kms}{km\,s$^{-1}$\xspace}
\newcommand{\Msun}{M$_\odot$\xspace}

\received{October 7, 2020}
\revised{October 15, 2020}
\accepted{November 17, 2020}

\submitjournal{ApJL}

\graphicspath{{./}{figures/}}

\usepackage{pgffor}
\usepackage[caption=false]{subfig}

\usepackage{amssymb}	
\usepackage{amstext}    
\usepackage{graphicx}

\begin{document}

\title{The First Integral Field Unit Spectroscopic View of Shocked Cluster Galaxies}

\correspondingauthor{Andra Stroe}
\email{andra.stroe@cfa.harvard.edu}

\author[0000-0001-8322-4162]{Andra Stroe}
\altaffiliation{Clay Fellow}
\affiliation{Center for Astrophysics \text{\textbar} Harvard \& Smithsonian, 60 Garden St., Cambridge, MA 02138, USA}

\author[0000-0002-0786-7307]{Maryam Hussaini}
\affiliation{University of Texas at Austin, Department of Astronomy, 2515 Speedway, Stop C1400 Austin, Texas 78712-1205, USA}
\affiliation{Center for Astrophysics \text{\textbar} Harvard \& Smithsonian, 60 Garden St., Cambridge, MA 02138, USA}

\author[0000-0003-2901-6842]{Bernd Husemann}
\affiliation{Max Planck Instit{\"u}t fur Astronomie, K{\"o}nigstuhl 17, Heidelberg, Germany}

\author[0000-0001-8823-4845]{David Sobral}
\affiliation{Department of Physics, Lancaster University, Lancaster LA1 4YB, UK}

\author[0000-0002-5445-5401]{Grant Tremblay}
\affiliation{Center for Astrophysics \text{\textbar} Harvard \& Smithsonian, 60 Garden St., Cambridge, MA 02138, USA}

\begin{abstract}
    Galaxy clusters grow by merging with other clusters, giving rise to Mpc-wide shock waves that travel at $1000-2500$\,\kms through the intra-cluster medium. To study the effects of merger shocks on the properties of cluster galaxies, we present the first spatially resolved spectroscopic view of 5 \Ha emitting galaxies located in the wake of shock fronts in the low redshift ($z\sim0.2$), massive ($\sim2\times10^{15}$\,\Msun), post-core passage merging cluster, CIZA J2242.8+5301 (nicknamed the `Sausage'). Our Gemini/GMOS-N integral field unit (IFU) observations, designed to capture \Ha and \NII emission, reveal the nebular gas distribution, kinematics and metallicities in the galaxies over $>16$\,kpc scales. While the galaxies show evidence for rotational support, the flux and velocity maps have complex features like tails and gas outflows aligned with the merger axis of the cluster. With gradients incompatible with inside-out disk growth, the metallicity maps are consistent with sustained star formation (SF) throughout and outside of the galactic disks. In combination with previous results, these pilot observations provide further evidence of a likely connection between cluster mergers and SF triggering in cluster galaxies, a potentially fundamental discovery revealing the interaction of galaxies with their environment.
\end{abstract}


\section{Introduction}\label{sec:intro}
The most extreme overdensities in the Universe evolve into the most massive ($\sim$10$^{15}$\,\Msun) gravitationally bound objects, galaxy clusters. Overdense environments heavily influence the evolution of galaxies: the densest parts of local, relaxed clusters are dominated by elliptical galaxies, devoid of ongoing star formation (SF) and the cold gas necessary for any future SF episode, while at lower densities the fraction of star-forming, gas-rich galaxies is larger than in the core \citep[e.g.][]{Dressler80, Solanes2001, Lewis2002, Tanaka2004, Mahajan2010}.

\begin{figure*}{}
    \centering
    \includegraphics[width=0.77\textwidth]{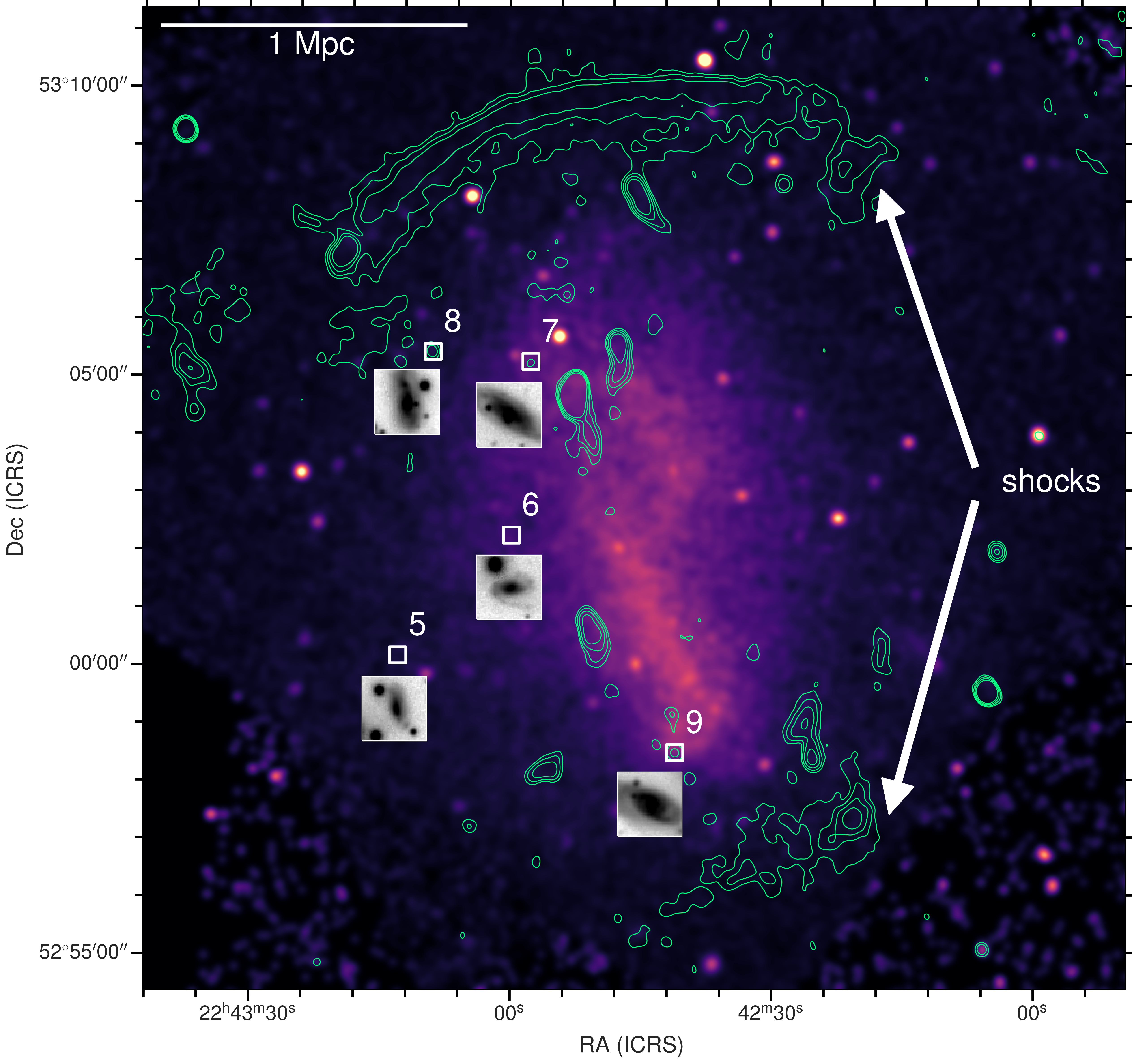}
    \caption{Our five shocked cluster galaxies in the `Sausage' cluster. Radio contours at 300\,MHz from the Giant Metrewave Radio Telescope \citep{Stroe2013} are drawn over a \textit{Chandra} X-ray image \citep{Ogrean2014}. The cluster has undergone a major merger in the plane of the sky, as evidenced by the elongated X-ray distribution. The large scale arc-like patches of radio emission located towards the north and south of the cluster trace large-scale shock waves induced by the merger 0.5--1\,Gyr ago. The white squares mark the positions of the targets followed-up with GMOS IFU spectroscopy. We display a $10^{\prime\prime}\times10^{\prime\prime}$ ($31.4$\,kpc$\times31.4$\,kpc) \textit{i}-band Subaru/Suprime-Cam \citep{Jee2015} zoom-in image on each target.}
    \label{fig:composite}
\end{figure*}

The role relaxed clusters have on galaxy evolution is well-established in the literature, but the picture is less clear for clusters undergoing a significant growth phase. Local, massive galaxy clusters gain most of their mass through mergers with less massive clusters, rather than infall of matter \citep{Muldrew2015}. In a simplified merger scenario, two clusters fall towards each other with speeds of thousands of $\rm{km}\,\rm{s}^{-1}$ and merge over the course of 1--2\,Gyr \citep[e.g.][]{Ricker2001}. As the two clusters pass through each other, significant energy is injected into the intracluster medium (ICM) in the form of large-scale bulk disturbances, fast-travelling shocks and cluster-wide turbulence. In the context of hierarchical structure formation, merging clusters are located at active nodes in the cosmic web and thus surrounded by an extensive network of filaments and smaller subclusters.

Merging clusters represent 30--50\% of the galaxy cluster population at $z<1$ \citep{Mann2012, Andrade-Santos2017, Rossetti2017} and of particular interest to the community as they present some surprising reversals of the typical environmental trends found in relaxed clusters. Studies contrasting statistical samples of relaxed and merging clusters, found that merging galaxy clusters have a higher density of emission-line, star-forming and blue galaxies, with higher specific SF rates (sSFR), stronger barred morphological features and large gas reservoirs and a higher fraction of active galactic nuclei (AGN) \citep{Miller2003, Cortese2004, Hwang2009, Hou2012, Jaffe2012, Jaffe2016, Sobral2015, Stroe2015a, Stroe2015, Stroe2017, Cairns2019, Yoon2019, Yoon2020}.

One of the most spectacular merging galaxy clusters is CIZA J2242.8+5301 ($z=0.188$, see Figure~\ref{fig:composite}), nicknamed the `Sausage'. The cluster hosts a unique overdensity of star-forming galaxies, over 25 times denser than the average cosmic volume at the cluster redshift and a SF rate (SFR) density $>$15 times above the level of typical star-forming galaxies at the `cosmic noon' ($z\sim$2--3), the peak of cosmic SF \citep{Stroe2014, Stroe2015a}. The cluster star-forming galaxies are massive, more metal-rich with lower electron densities compared to field galaxies and show evidence for outflows, driven either by supernovae or AGN \citep{Sobral2015, Stroe2015a}. Moreover, the `Sausage' displays evidence for sustained SF over time scales of 500\,Myr, as well as large neutral gas reservoirs to fuel future SF episodes \citep{Stroe2015}.

To understand the evolution of SF in cluster galaxies, we must take into account the extraordinary merger history of the `Sausage' cluster. The `Sausage' cluster went through a massive merger $<$1\,Gyr ago, along the north--south direction, in the plane of the sky, between two progenitors, each $\sim10^{15}$\,\Msun with a relative speed of 2000--2500\,\kms \citep{vanWeeren2011, Stroe2014b, Jee2015}. Two symmetric, fast-moving ($\sim$1000--2500\,\kms) shocks were produced as the two subclusters passed through each other 0.5--1\,Gyr ago. The shock waves propagated through the ICM along the merger axis and are revealed as arc-like, Mpc-wide patches of diffuse radio emission at the cluster outskirts \citep[Figure \ref{fig:composite}, e.g.][]{Stroe2013}.

\begin{deluxetable*}{cccDDcccc}[htb!]
    \tablecaption{Basic properties of the galaxies, including coordinates, redshift, stellar mass, \Ha flux, \textit{i} band AB magnitude, an estimate of the Hubble morphological classification and the classification as SF or AGN from the 1D spectroscopy \citep[measurements from][]{Sobral2015}. \label{tab:prop}}
    \tablewidth{0pt}
    \tablehead{
        \colhead{Target}  &  \colhead{R.A.}  & \colhead{Decl.}                                 &  \multicolumn2c{z} & \multicolumn2c{M$_{\star}$} & \colhead{F$_{\mathrm{H}\alpha}$\tablenotemark{a}} & \colhead{$i$\tablenotemark{b}}  & \colhead{Morphology} & \colhead{Classification} \\
        & \colhead{hh\,:\,mm\,:\,ss.ss} & \colhead{$^\circ\,:\,{^\prime}\,:\,{^{\prime\prime}}$}  & \multicolumn2c{} & \multicolumn2c{($10^9$\,\Msun)} & \colhead{($10^{-15}$\,erg\,s$^{-1}$\,cm$^{-2}$)} & \colhead{(mag)}  & &
    }
    \decimals
    \startdata
    Sausage 5 & 22:43:12.90 & +53:00:10.08   & 0.182475  & $3.9\pm1.6$ & 1.78 &   18.94     &  SBa  & SF     \\
    Sausage 6 & 22:42:59.85 & +53:02:14.57  & 0.1836   & $2.6\pm1.7$ & 1.88 &   18.70      &  SBc & SF    \\
    Sausage 7 & 22:42:57.63 &  +53:05:14.75  & 0.18315  & $33.5\pm10.6$ & 2.13 &   17.72      &  Sb  & SF + outflows    \\
    Sausage 8 & 22:43:08.88 &  +53:05:25.04  & 0.1843   & $42.8\pm13.2$ & 4.06 &   17.42      &   SBa & AGN    \\
    Sausage 9 & 22:42:41.07 &  +52:58:28.67  & 0.18394    & $45.1\pm13.8$ & 5.14 &   17.35      &  Sb  & AGN + outflows    \\
    \enddata
    \tablenotetext{a}{Errors typically $<10$\%.} \tablenotetext{b}{Errors $<0.01$\,mag.}
\end{deluxetable*}

To explain the unusual nature of the star-forming galaxies in the `Sausage' cluster, \citet{Stroe2015a}, in agreement with models from \citet{Roediger2014} and \citet{Ebeling2019}, speculate that the travelling shocks pass through the gas-rich cluster galaxies, disrupt the gas to trigger SF and fuel an AGN. Similarly to infalling galaxies experiencing ICM ram pressure \citep[e.g.][]{Gunn1972}, the shock interaction model predicts a disruption of the gaseous disk in galaxies located in the wake of the shock fronts and the presence of tails, knots or filaments of ionized gas aligned with the merger axis \citep[e.g.][]{Ebeling2019}.

What causes the surprising SFR in the `Sausage' cluster galaxies? In this paper, we put the shock-induced SF model to the test. We present Gemini GMOS-N IFU spectroscopic observations of five, \Ha-selected, star-forming, active galaxies within the `Sausage' cluster (see Figure~\ref{fig:composite}, Table~\ref{tab:prop}), which unveil the \Ha and \NII (6585\,\AA{}) gas dynamics, the detailed \Ha morphologies and resolved metallicities and provide direct proof on whether the ionized gas regions in cluster galaxies are disrupted by the passage of the shock waves.

We assume a flat $\Lambda$CDM cosmology, with $H_0 = 70.0$\,\kms\,Mpc$^{-1}$, $\Omega_{m}=0.3$ and $\Omega_{\Lambda}=0.7$. At the redshift of the cluster, 1$^{\prime\prime}=3.142$\,kpc. 

\section{Targets, Observations and Data Reduction}
\label{sec:obs}

\subsection{Target selection}
Our targets are drawn from our SFR limited narrow-band survey, which uniformly selects \Ha emitters in the `Sausage' cluster and the cosmic web around it \citep{Stroe2015a}. For the present study, we focus on five galaxies (see Figure~\ref{fig:composite}, Figure~\ref{fig:spectra} and Table~\ref{tab:prop}), confirmed as cluster members in our follow-up spectroscopic survey \citep{Sobral2015}. For the IFU follow-up, the targets were chosen to be massive ($\gtrsim2\times10^{9}$\,\Msun), bright, with \Ha fluxes of $\gtrsim10^{-15}$\,erg\,s$^{-1}$\,cm$^{-1}$, and \textit{i} band magnitudes between 17 and 20 mag (AB), and spatially extended over $\gtrsim5^{\prime\prime}$ (equivalent to $\gtrsim16$\,kpc). The galaxies are located in post-shock regions, traversed by shock waves as recently as 500\,Myr. Three galaxies are powered primarily by SF, while two have significant contributions from AGN, while still presenting morphologies consistent with spiral structure. Based on the 1D spectroscopy, two galaxies have evidence for outflows, powered by SF or AGN \citep{Sobral2015}. As such, the sample enables us to test the effect of the cluster merger and the shock waves on the triggering of SF and black hole activity.

\subsection{Observations and Data Reduction}
We observed five galaxies with the Gemini Multi-Object Spectrograph (GMOS\footnote{\url{http://www.gemini.edu/instrumentation/current-instruments/gmos}}) in the IFU mode (Gemini program GN-2018B-Q-318, PI Stroe). With the 2-slit configuration, the observations covered a $5^{\prime\prime}\times7^{\prime\prime}$ field-of-view (FOV) centered on the galaxy, and a $5^{\prime\prime}\times3.5^{\prime\prime}$ sky area, offset by $1^\prime$ from the target. We obtained six exposures of 813\,s for each target in queue mode observing, taking advantage of the poorer observing conditions at Mauna Kea, with gray moon, cloudy weather and image quality varying between 0.50--0.75$^{\prime\prime}$ at zenith, as measured on a point source during acquisition. We used the R150 grating in combination with the GG455 filter and a central wavelength of 7300\,\AA{} and 7600\,\AA{} to cover the wavelength gaps. This setup results in a contiguous 5100--9900\,\AA{} coverage at $\sim300$\,\kms resolution.

\subsection{Data Reduction}
The data were reduced with the Py3D data reduction package for fibre-fed IFU spectrographs initially developed for the CALIFA survey \citep{Husemann:2013}. It has already been successfully applied to similar GMOS IFU data beforehand \citep[e.g.][]{Husemann:2016}. We perform basic reduction steps on individual exposures which include bias subtraction, cosmic ray cleaning using PyCosmic \citep{Husemann:2012}, fiber identification, fiber tracing, stray light subtraction, optimal spectral extraction, wavelength calibration based on CuAr arc lamps and fibre flat-fielding based on a twilight observation. The standard star Wolf 1346, used as flux calibrator, is calibrated in the same way as the data to determine the sensitivity curve for the given instrumental setup. The individual exposures were flux calibrated before a mean sky background spectrum was constructed from the dedicated offset sky fiber and subsequently subtracted from all fiber spectra in the target FoV. A final data cube is reconstructed by drizzling \citep{Fruchter:2002} all fibers into a regular grid of squared pixels with a $0.2^{\prime\prime}$ sampling. With emission lines masked, we collapsed the cube between 7400--8150\,\AA{} to obtain a continuum image, which was used to refine the astrometry (within $0.2^{\prime\prime}$) and precisely align the IFU for each galaxy to its \textit{i}-band image. 

\begin{figure*}[htbp!]
    \centering
    \includegraphics[width=\textwidth]{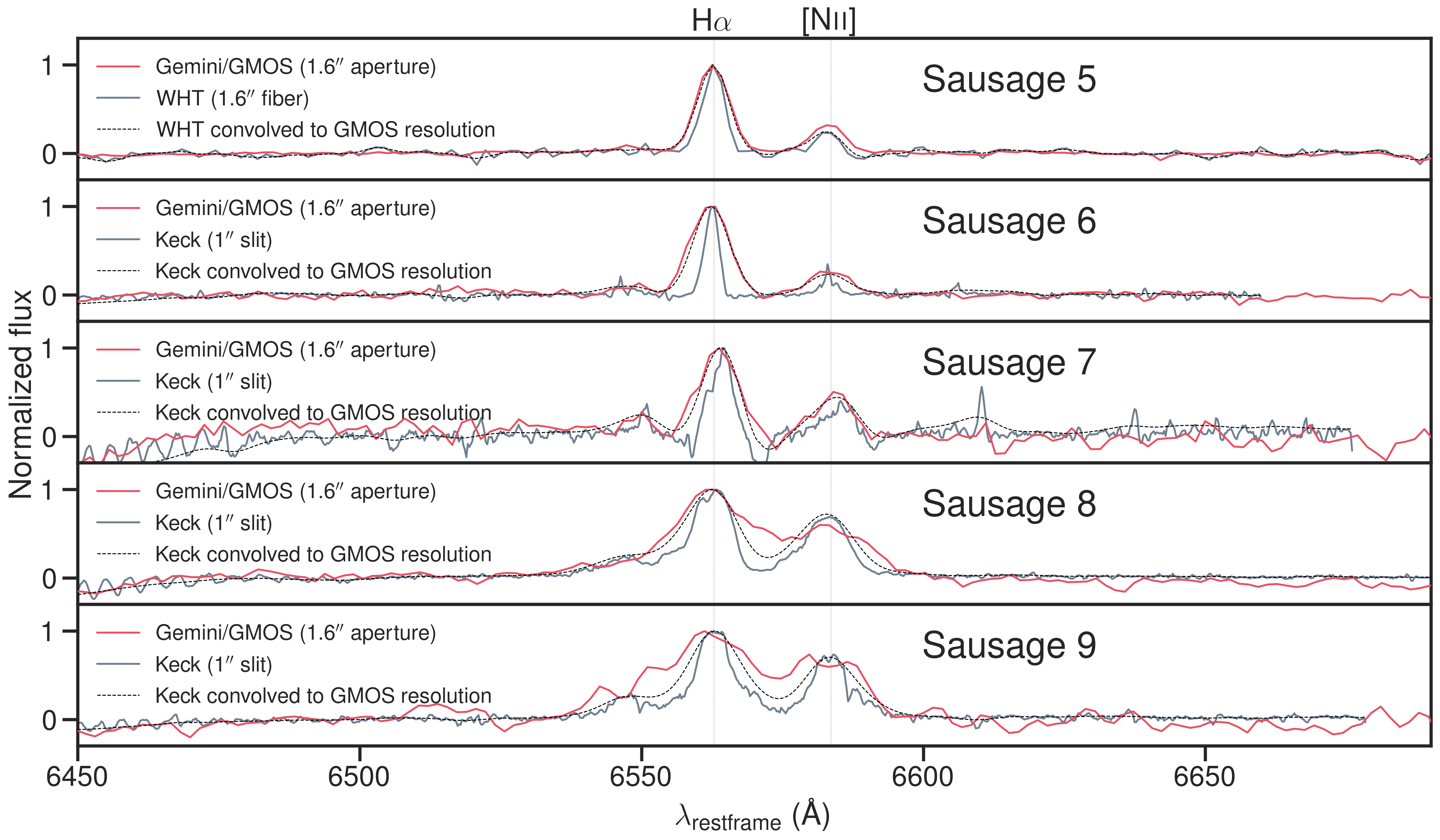}
    \caption{Aperture (1.6$^{\prime\prime}$) spectra extracted from the Gemini/GMOS IFU observations are in excellent agreement with the 1D spectra from our 1.6$^{\prime\prime}$ fiber and 1$^{\prime\prime}$ slit observations \citep{Sobral2015}. The spectra are continuum subtracted and normalized to the \Ha peak.}
    \label{fig:spectra}
\end{figure*}

We employ a 2D Gaussian filter with a 1.2 pixel standard deviation to spatially smooth the data. Overall, we find excellent agreement between spectra extracted in $1.6^{\prime\prime}$ apertures from the IFU observations and our slit and fiber observations from \citet{Sobral2015} (see Figure~\ref{fig:spectra}). \Ha and \NII are detected at high signal to noise ratio (S/N) in all the galaxies, while faint continuum emission and \Hb, \OIII and \SII are detected at lower S/N only in some of the galaxies. For the rest of the paper, we focus solely on the analysis of the \Ha and \NII emission lines in line with our main science goals.

\section{Analysis}\label{sec:analysis}

We use the \gleam\footnote{\url{https://github.com/multiwavelength/gleam}} Python package \citep[\textbf{G}alaxy \textbf{L}ine \textbf{E}mission \& \textbf{A}bsorption \textbf{M}odelling,][]{stroe_gleam} to jointly fit the \Ha and \NII emission lines and the continuum emission for each spaxel in the smoothed GMOS-N cubes. A window 140\,\AA{} wide around \Ha and \NII is modelled with a constant plus two Gaussian models \footnote{\NII (6550\,\AA) contribution is negligible}, which are all free parameters in the fit. The redshift measured from the 1D spectroscopy \citep{Sobral2015} is used as starting solution for the Gaussian center. The positions of \Ha and \NII was allowed to independently vary within 9\,\AA{} (or $\pm350$\,\kms) around the wavelength predicted by the systemic redshift. We require a S/N of 3 for emission line detections. 

\begin{figure*}[htbp!]
    \centering
    \includegraphics[width=\textwidth]{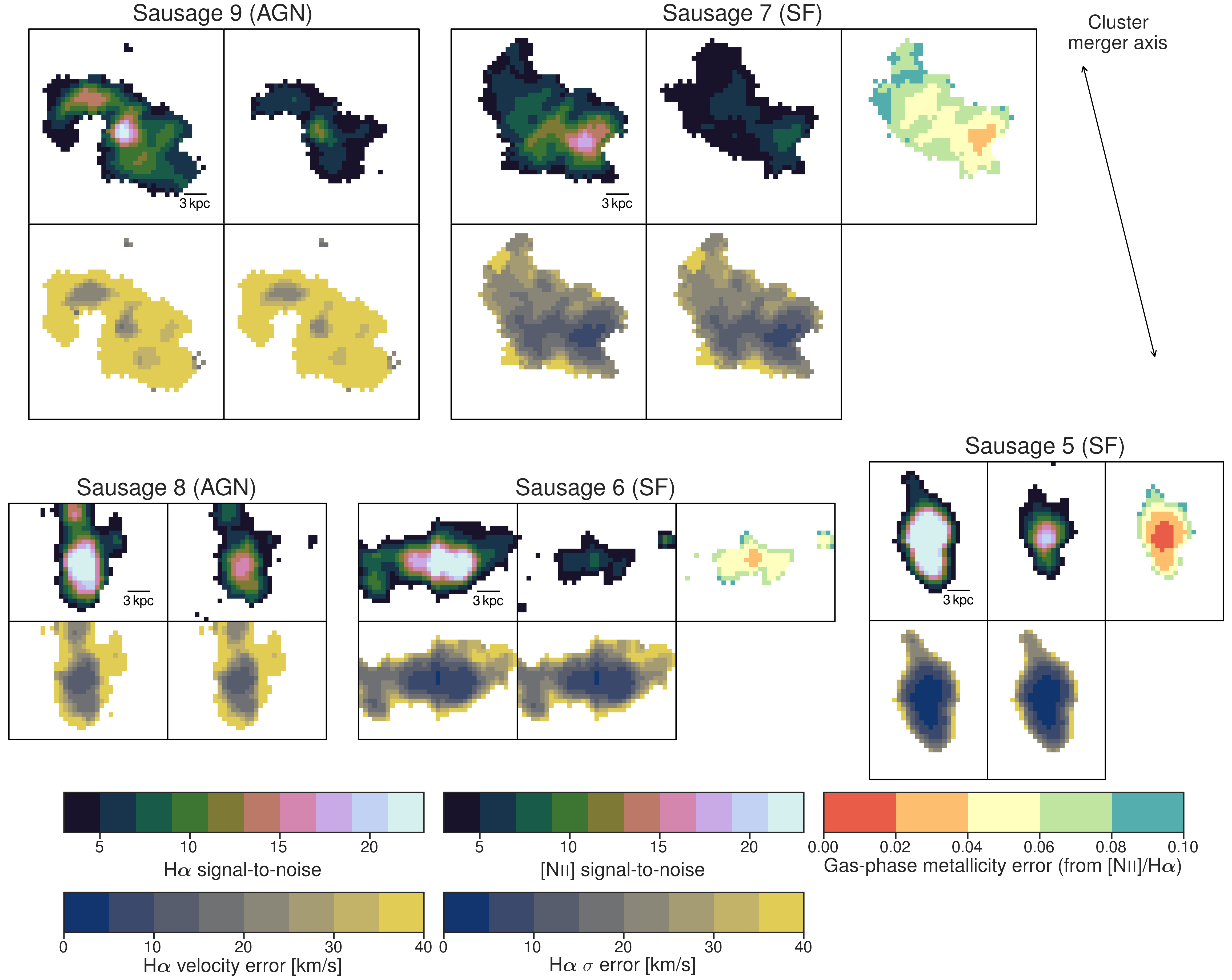}
    \caption{Error maps corresponding to the gallery of galaxies presented in Figure~{fig:gallery}. For each galaxy, clockwise from the top left, we show the \Ha S/N map, the \NII S/N map, the \Ha velocity dispersion uncertainty map and \Ha velocity uncertainty map. For star forming galaxies, we also show the metallicity error map. In the \NII flux and metallicity maps we show the border of the \Ha emission in light gray.}
    \label{fig:errors}
\end{figure*}

We build flux, velocity and dispersion maps from the continuum subtracted fluxes and line velocities, as reported from the Gaussian fits (Figure~\ref{fig:gallery}) and associated S/N and error maps (Figure~\ref{fig:errors}). The minimum dispersion measurable (120\,\kms) is limited by the instrumental resolution. For SF-dominated galaxies and in regions where both \NII and \Ha are detected at S/N$>3$, we use the \NII/\Ha line ratio to derive spatially resolved metallicity maps (Figure~\ref{fig:gallery}). Using the calibration from \citet{Pettini2004}, we convert the \NII/\Ha ratio to metallicity (oxygen abundance): $12 + \log_{10}(\mathrm{O/H}) =  8.9 + 0.57\log_{10}(\mathrm{[N}\textsc{ii}]/\mathrm{H}\alpha)$. 

\section{Flux, dynamics and metallicity maps}\label{sec:results}

\begin{figure*}[htbp!]
    \centering
    \includegraphics[width=\textwidth]{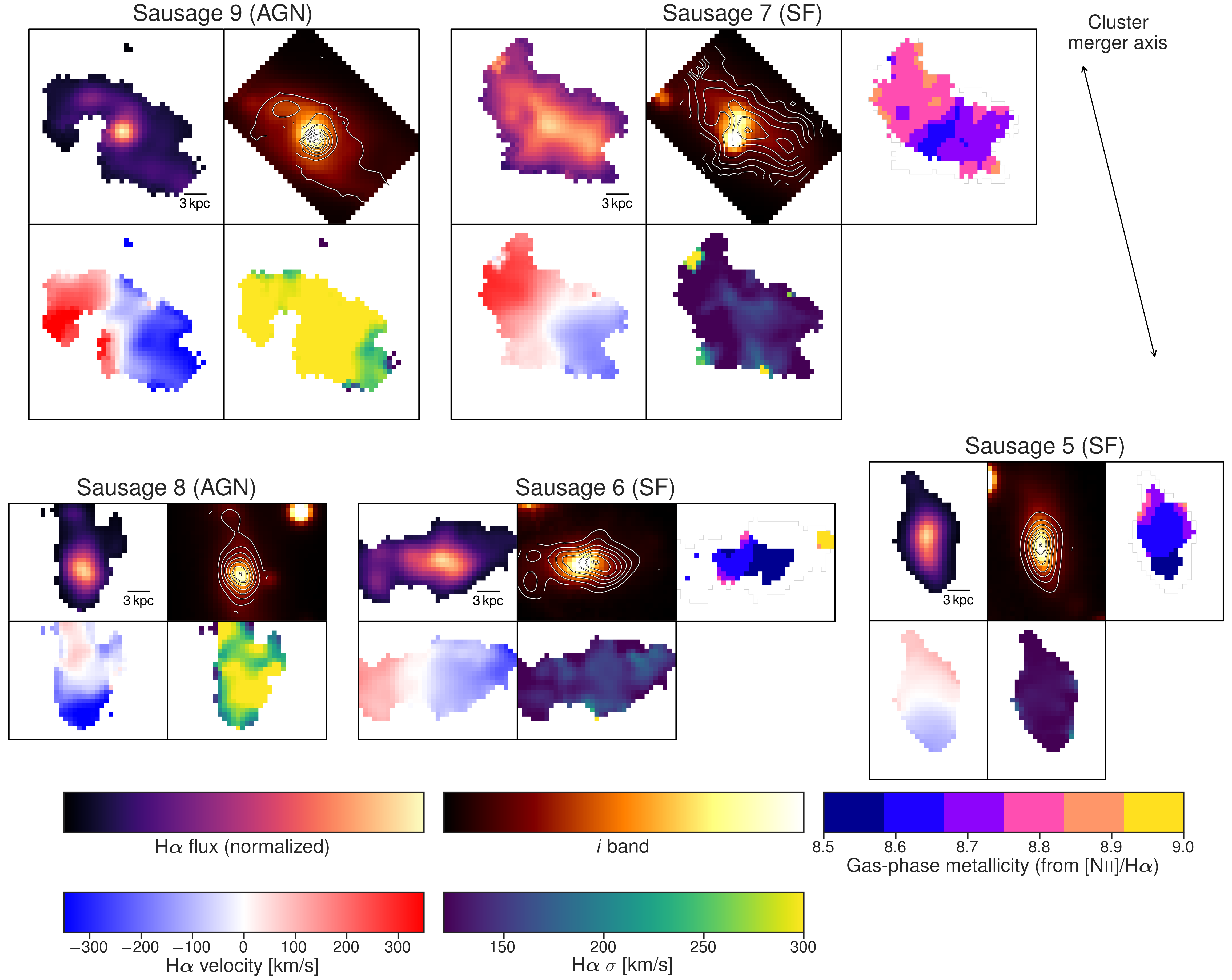}
    \caption{A gallery of `shocked' galaxies in the `Sausage' cluster merger. For each galaxy, clockwise from the top left, we show the \Ha flux map, \textit{i}-band Subaru/Suprime-Cam image with \Ha flux contours, \Ha dispersion and velocity map. For SF-dominated galaxies, we also show the metallicity map, with the \Ha detection border shown in light gray. All galaxies show evidence for morphologies and kinematics disturbed in the north-south direction. The alignment with the merger axis of the cluster suggests a tight connection between the merger event and the triggering of SF. Corresponding S/N and error maps can be found in Figure~\ref{fig:errors}.}
    \label{fig:gallery}
\end{figure*}

Figure~\ref{fig:gallery} shows a gallery of our five targets, unveiling their \Ha flux, \Ha velocity and dispersion, together with an \textit{i} band optical image. We also show a metallicity map for the star forming galaxies. We robustly detect dynamics of nebular \Ha and \NII emission extended over 16\,kpc in all five galaxies, with evidence for disturbed morphologies, including tails of ionized gas offset from the stellar disk. The metallicity maps show diverse distributions.

Sausage 5, 6, 7 are firmly classified as SF spiral galaxies based on line ratios in their 1D spectra \citep{Sobral2015}. Further evidence to support this scenario comes from the IFU data, where the emission is consistent with photoionization throughout the galaxies, considering the small ratios between the \SII doublet (which in many cases is not detected) and \Ha \citep{Kewley2006}. All three galaxies have strong \Ha velocity gradients with comparatively low dispersions, confirming their rotating nature. All three galaxies have bright knots of ionized gas emission with high relative velocities of $\pm300$\,\kms. Generally, the \Ha flux extensions follow the motion of the disk, but with larger amplitudes in velocity. In Sausage 5, the peak of the \Ha emission is offset in the north direction by about $0.2^{\prime\prime}$ ($\sim0.6$\,kpc). The ionized nebular gas in Sausage 5 has an asymmetric rotation pattern with higher amplitude in velocity on the approaching side ($\sim-100$\,\kms with 7--8\,\kms uncertainty per pixel) and velocities of up to $70\pm23$\,\kms at the northern tip, an offset flux peak towards the north-west from the stellar disk. The tails and spurs of ionized gas are detected at $\textrm{S/N}\sim$4--10 per pixel across the features. For example, in Sausage 6, the \Ha emission peak, embedded in a region of metal-poor gas ($8.5\pm0.05$ per pixel) is offset $0.6^{\prime\prime}$ ($\sim1.9$\,kpc) north-west from the peak of the stellar emission, followed by a spur of extremely metal-rich gas ($9.0\pm0.05$ per pixel) towards the north-west of the galaxy (offset $3.7^{\prime\prime}$ west and $1.2^{\prime\prime}$ north from the stellar disk center). In Sausage 7, the ionized gas maps look remarkably different from the stellar distribution. The bright nucleus surrounded by a well-defined spiral structure is not reflected in the complex, clumpy \Ha gas, whose bow-like distribution is offset north from the stellar disk. The peak of the \Ha emission is also offset $0.5^{\prime\prime}$ ($\sim1.6$\,kpc) from the peak of the stellar light. The metallicity has a strong 0.2\,dex gradient along the disk of the galaxy, with elevated values on the side closest to the northern cluster-scale shock front.

While the emission line budget in both Sausage 8 and 9 is dominated by AGN contribution \citep{Sobral2015}, the resolved IFU observations reveal a more complex picture. With the fastest \Ha rotational velocities but also the largest gas dispersions of the sample (over $\sim400$\,\kms), Sausage 9 is a classical Seyfert 1 type source, with a bright nucleus dominated by AGN emission, broad emission lines and a pronounced spiral arm pattern powered by SF, recovered in both the \textit{i} band and the ionized gas maps. The peak of \Ha emission is offset south in Sausage 9, by $\sim0.9$\,kpc with respect to the optical nucleus. Sausage 8 shows two tails of \Ha emission distinct from the general disk rotation pattern: one tail of redshifted gas and a spur of blueshifted \Ha emission towards the north-west. 

\section{Discussion}
We explore the role of the merging cluster environment in triggering sustained SF in five massive, gas-rich, main-sequence galaxies in the post-shock region within the `Sausage' cluster. Our IFU observations reveal morphological and kinematical disturbances in the nebular gas, generally aligned with the merger axis of the cluster. Our main aim is to disentangle whether the high-significance tails, spurs and knots are caused by infall and interaction with the ICM, by galaxy-galaxy mergers/interactions or by a cluster-wide process, such as a merger-induced shock.

\subsection{Infalling galaxies?}
The majority of isolated and undisturbed galaxies have regular kinematic maps and strong negative gas-phase metallicity gradients \citep[e.g.][]{Poetrodjojo2018}, explained by an inside-out disk formation model, where the central metal-rich region has been undergoing sustained SF for longer than the metal-poor gas at the outskirts of the galaxy \citep[e.g.][]{Pilkington2012}. For star-forming cluster galaxies, we might expect a large fraction of disturbed \Ha kinematics and morphologies due to the interaction with the ICM during their infall and gravitational perturbations \citep{Cortese2007}. However, in their large sample of \Ha-selected galaxies, \citet{Tiley2020} find a variety of \Ha morphologies, including regular, centrally-peaked distributions with disk-like velocity maps and irregular distributions in both field and cluster environments, as galaxies undergoing significant quenching would not be included in their \Ha-selected sample. 

A particularly interesting class to compare with are jellyfish, galaxies infalling into clusters which exhibit gaseous tails with bright star-forming knots caused by ram pressure. Despite the extreme interaction with the ICM, jellyfish galaxies display strong negative gas-phase metallicities from the center towards the stripped tails, consistent with an inside-out formation in isolation, followed by a outside-in removal of gas upon infall into the cluster \citep[e.g.][]{Bellhouse2019, Franchetto2020}. Our galaxies show offset \Ha flux peaks, extended tails of nebular emission and bow-like and asymmetric velocity maps reminiscent of those seen in jellyfish galaxies. However, the tantalizing alignment of these features with the merger axis is broadly incompatible with galaxies infalling into the cluster radially and, unlike jellyfish galaxies, we actually find consistent evidence against radial metallicity gradients. 

{We can draw comparisons between detailed studies of infalling galaxies in local clusters. For example, \citet{Chemin2006} conducted a detailed kinematic analysis of 30 typical spiral galaxies in the $1.2\times10^{15}$\,M$_{\odot}$ Virgo cluster, located at a distance of just $\sim16.5$\,Mpc. The bulk of their sample is located outside the core of the cluster, specifically at relative velocities and radii larger than our sample (see phase-space diagram in Figure~\ref{fig:phase}). \citet{Chemin2006} find evidence for disturbed \Ha kinematics, but the offsets between the peak of the stellar light and the \Ha are of the order of $0.2-0.4$\,kpc, smaller than what we observe in our galaxies ($0.6-1.9$\,kpc). Outside of its core, the Coma cluster ($7\times10^{14}$\,M$_{\odot}$, located at $\sim100$\,Mpc) is dominated by disturbed galaxies with tails of \Ha emission \citep{Gavazzi2018}. Three Coma galaxies are located to the cluster center as close as our galaxies Sausage 6-9 (Figure~\ref{fig:phase}). However, these three galaxies are undergoing extreme ram pressure. With only one galaxy powered by SF and two by AGN, there is little to no \Ha within the stellar disk and the bulk of \Ha is found outside the galaxies, streaming out in long tails \citep{Yagi2010}. Observations of both Coma and Virgo indicate that infalling galaxies can present tails of \Ha emission and perturbed kinematics, but the \Ha gas is removed from the outside in, while the peak of the emission, close to the nucleus of the galaxies does not get displaced.

Considering that our galaxies are deeply embedded in the hottest parts of the ICM (see Figure~\ref{fig:phase}) and present disturbed morphologies and kinematics throughout and outside the stellar disk, it is improbable that the nebular gas features are caused by ram pressure in infalling galaxies. The orientation of the gas tails imply infall pathways for galaxies 6, 7, 8 and 9 which would cross the densest, hottest parts of the ICM. Under the assumption of infall, the gas reservoirs in these galaxies would be almost completely depleted, as evidenced by detailed analyses of the Coma and Virgo clusters \citep{Chemin2006, Yagi2010,Gavazzi2018}.

\begin{figure}[htbp!]
    \centering
    \includegraphics[width=0.5\textwidth]{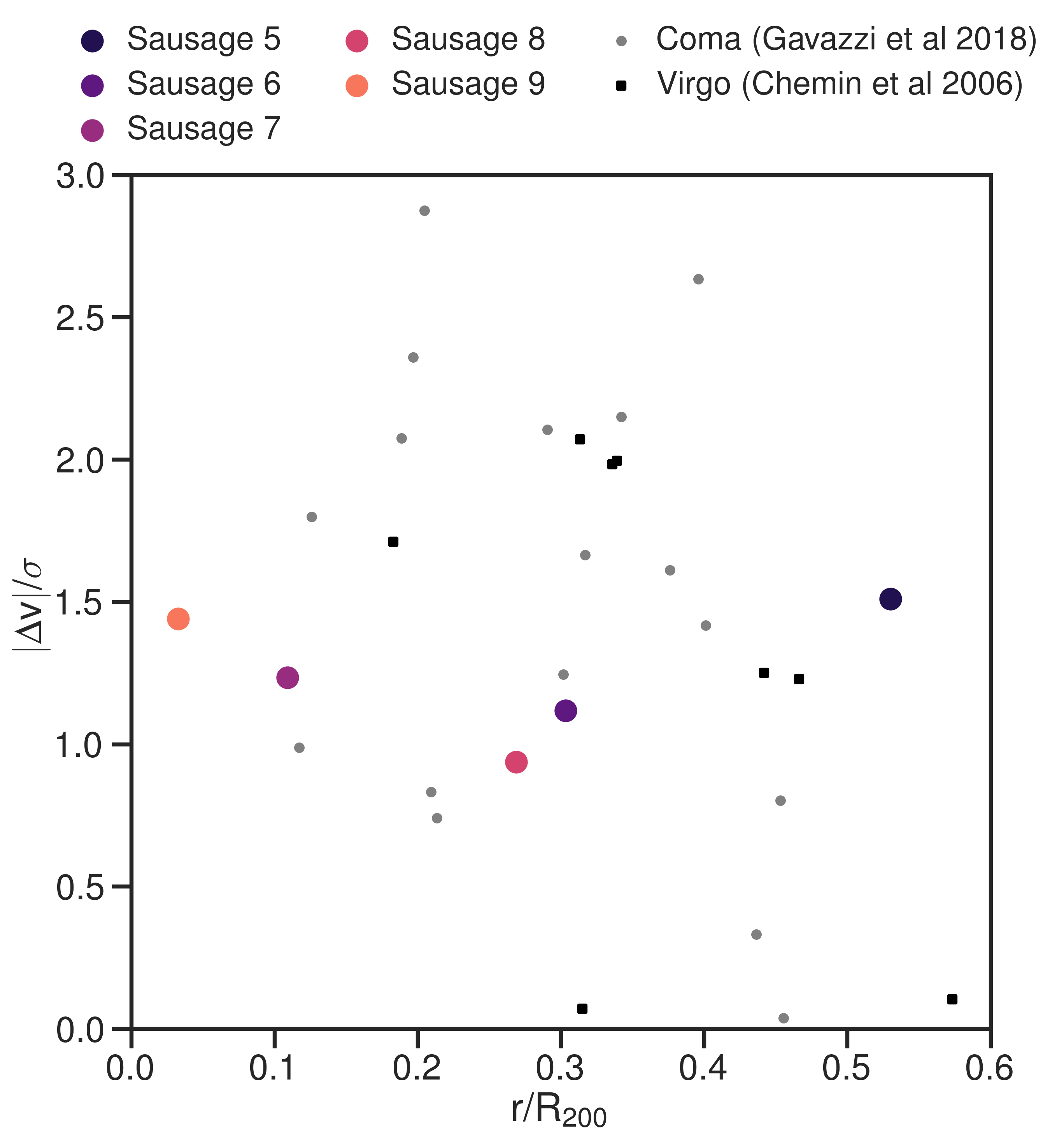}
    \caption{Phase-space diagram for the cores of the Sausage, Virgo and Coma clusters, highlighting kinematically disturbed \Ha galaxies. We show our five galaxies with respect to the properties of the closest subcluster \citep[as derived from weak lensing and dynamics in][]{Jee2015}, together with data from \citet{Chemin2006} on the Virgo cluster and \citet{Gavazzi2018} for the Coma cluster. Our galaxies are embedded deep within the ICM. Unlike galaxies at similar cluster-centric radii and relative velocities in Coma and Virgo which are almost completely devoid of \Ha within the disk, the Sausage galaxies show \Ha throughout the stellar disk, with large offsets between the peak of \Ha and the stellar light.}
    \label{fig:phase}
\end{figure}

\subsection{Interacting galaxies?}
The kinematic disturbances seen in our sample are reminiscent of those seen in interacting or merging galaxies \citep[e.g.][]{Torres2014}. Additionally, metallicity gradients can be much shallower in lower mass galaxies and galaxies that are disturbed, for example by a tidal interaction or a merger with another galaxy which can funnel pristine, metal-poor gas towards the core of the galaxy \citep[see review by][]{Kewley2019}. The interaction scenario explains some, but not all of the observations: our galaxies maintain regular kinematics within the bulk of the stellar light and all three star-forming galaxies have metallicity gradients across the stellar disk. Since galaxy-galaxy interactions are most common in low-mass clusters and group-like environments, it is highly unlikely that all five galaxies are undergoing mergers in an extremely massive, $2\times10^{15}$\,\Msun cluster such as the `Sausage'.

\subsection{SF induced by cluster merger?} 
Fast-travelling, relatively low-Mach number ($M\sim$1--4) shocks, such as those produced in cluster mergers \citep{Roediger2014}, possibly compounded with the time-dependent tidal fields of merging clusters \citep[e.g.][]{Bekki1999} are expected to result in elevated pressure which compresses the interstellar medium, not only at the outskirts, but also deep within galaxies, and results in galaxy-wide SF sustained on 100--500\,Myr timescales. The features we observe in the flux, velocity and metallicity maps of our galaxies can be ascribed to this scenario. Unlike a galaxy-galaxy merger, the shock front would not `destroy' the disk structure of the galaxy, but cause ionized tails broadly aligned with the merger direction and propagation direction of the shock. This is reflected in our nebular gas kinematic maps, which have disk-like kinematics with \Ha and \NII tails, spurs and offsets in the general north-south direction, despite the range of position angles the galaxies span on the sky. The metallicity maps reveal SF trends inconsistent with a simple inside-out disk formation scenario. Located only 500\,kpc away from the northern shock front, Sausage 7 was traversed by the shock less than 200\,Myr ago and is thus expected to show strong evidence for shock interaction. Indeed, the source's elevated metallicities support a scenario of elevated SF activity throughout the galaxy. The  star-forming sources all show metallicity gradients across the galaxy disk which broadly follow the distribution of the disrupted offset nebular gas, suggesting a plausible causal connection between the shock disruption and the triggering of SF. Overall, SF is quenched in clusters with significant populations of galaxies disturbed by ram pressure or tidal fields. By contrast, the Sausage cluster presents ample evidence of a significant enhancement of SF \citep{Stroe2015,Stroe2017}. If ram pressure or tidal disruption alone cause the features in our velocity, metallicity and gas maps, similar SF enhancements should be visible in other clusters. When taking all evidence into account, both presented in the present paper and in previous analyses, the simplest scenario that explains all the data suggests a likely causal connection between the merger of the cluster and the features observed in our data.

\section{Conclusion}\label{sec:conclusion}

We presented the first resolved IFU spectroscopic observations of five \Ha-selected main-sequence galaxies in the low-redshift ($z\sim0.2$), massive ($\sim2\times 10^{15}$\,\Msun), `Sausage' merging cluster, which displays a surprising reversal of the typical environmental trends observed in $z<1$ clusters. The five galaxies have disk-like \Ha morphologies and kinematics, with evidence of disturbed \Ha and \NII tails and spurs aligned with the merger axis of the cluster. Metallicity gradients are consistent with SF triggered throughout the galaxies. These observations possibly present the most direct evidence for SF induced by the merger of massive galaxy clusters and their associated large-scale shock waves, especially when combined with previous results of elevated SF in the Sausage cluster.

The pilot observations shown here demonstrate that leaps in our understanding of galaxy cluster physics are achievable with IFUs, even with modest telescope time investments. Future studies of statistically significant samples will disentangle shock, merger and ram pressure contributions in triggering SF across a range of local densities and stellar masses.

\section{Acknowledgements}

We thank the referee for their excellent comments that have improved the paper. We thank Adrian Bittner, Jorryt Matthee and Rebecca Nevin for useful discussions. Andra Stroe gratefully acknowledges support of a Clay Fellowship. Maryam Hussaini acknowledges the Smithsonian Astrophysical Observatory REU program, which is funded in part by the National Science Foundation REU and Department of Defense ASSURE programs under NSF Grant no.\ AST-1852268, and by the Smithsonian Institution. Bernd Husemann acknowledges financial support by the DFG grant GE625/17-1 and DLR grant 50OR1911. We thank Matthew Ashby and Jonathan McDowell for comments on an early draft. Based on observations obtained at the international Gemini Observatory, a program of NSF's NOIRLab, which is managed by the Association of Universities for Research in Astronomy (AURA) under a cooperative agreement with the National Science Foundation, on behalf of the Gemini Observatory partnership: the National Science Foundation (United States), National Research Council (Canada), Agencia Nacional de Investigaci\'{o}n y Desarrollo (Chile), Ministerio de Ciencia, Tecnolog\'{i}a e Innovaci\'{o}n (Argentina), Minist\'{e}rio da Ci\^{e}ncia, Tecnologia, Inova\c{c}\~{o}es e Comunica\c{c}\~{o}es (Brazil), and Korea Astronomy and Space Science Institute (Republic of Korea). Based in part on data collected at Subaru Telescope, which is operated by the National Astronomical Observatory of Japan. The authors wish to recognize and acknowledge the very significant cultural role and reverence that the summit of Maunakea has always had within the indigenous Hawaiian community. We are most fortunate to have the opportunity to conduct observations from this mountain.

\vspace{5mm}
\facilities{Gemini:Gillett (GMOS-N), Subaru (Suprime-Cam), CXO (ACIS-I), GMRT}

\software{
    \textbf{gleam} \citep{stroe_gleam},
    Astropy \citep{2013A&A...558A..33A},
    APLpy \citep{aplpy},
    DS9 \citep{ds9},
    QFitsView
}

\bibliography{AStroe_Shocked_Galaxies}{}
\bibliographystyle{aasjournal}
    
\end{document}